\documentclass[twocolumn,runningheads,natbib]{svjour2}
\smartqed  
\usepackage{graphicx}
\newcommand\psr{AX J1845.0--0258}
\newcommand\cxou{CXOU J184454.6--025653}
\newcommand\axj{AX J184453--025640}
\newcommand\xte{XTE J1810--197}
\newcommand\units{erg~s$^{-1}$~cm$^{-2}$}
\journalname{Astrophysics and Space Science}
\begin{document}

\title{\textit{Chandra} Monitoring of the Candidate Anomalous X-ray
  Pulsar \psr 
}


\author{Cindy R. Tam \and Victoria M. Kaspi \and Bryan M. Gaensler
  \and	Eric V. Gotthelf  
}


\institute{C. R. Tam \and V. M. Kaspi 
          \at
              Department of Physics, Rutherford Physics Building,
              McGill Univeristy, 3600 University Street, Montreal, QC
              H3A 2T8, Canada. \\
              \email{tamc@physics.mcgill.ca}           
           \and
           B. M. Gaensler
	   \at
              Harvard-Smithsonian Center for Astrophysics, 60 Garden
              Street, Cambridge, MA 02138, USA.  \emph{Present
	      address:} School of Physics, University of Sydney, NSW
              2006, Australia.
	   \and
	   E. V. Gotthelf 
	   \at
	      Columbia Astrophysics Laboratory, Columbia University,
              550 West 120th Street, New York, NY 10027-6601, USA.
}

\date{Received: date / Accepted: date}

\maketitle

\begin{abstract}
The population of clearly identified anomalous X-ray pulsars has
recently grown to seven, however, one candidate anomalous X-ray pulsar
(AXP) still eludes
re-confirmation.  Here, we present a set of seven
\textit{Chandra} ACIS-S observations of the transient pulsar \psr,
obtained during 2003.
Our observations reveal a faint X-ray point source within 
the \textit{ASCA} error circle of \psr's discovery, which we
designate \cxou\ and tentatively 
identify as the quiescent AXP.  Its spectrum is well described by an
absorbed single-component blackbody ($kT\sim2.0$~keV) or power law
($\Gamma\sim1.0$) that is steady in flux on timescales of at least
months, but fainter than \psr\ was during its 1993 period of X-ray
enhancement by at least a factor of 13.  Compared to the outburst
spectrum of \psr, \cxou\ is considerably harder: if
truly the counterpart, then its spectral behaviour is contrary to that
seen in the established transient AXP \xte, which softened from
$kT\sim0.67$~keV to $\sim0.18$~keV in quiescence.  This
unexpected result prompts us to examine the possibility that we have
observed an unrelated source, and we discuss the implications for
AXPs, and magnetars in general.

\end{abstract}

\section{Introduction to \psr}
\label{intro}

The 6.97-s X-ray pulsar \psr\ was discovered serendipitously in
archival \textit{ASCA} observations from 1993 \citep{gv98,tkk+98}
during a period of apparent outburst.  Its slow spin period, soft
spectrum ($kT\sim0.64$~keV) and positional coincidence with the newly
discovered supernova remnant G29.6+0.1 \citep{ggv99} were suggestive
of the small but growing class of anomalous X-ray pulsars (AXPs, see
review by V. Kaspi, this volume\footnote{For a summary of AXP
properties, see the online 
catalog http://www.physics.mcgill.ca/$\sim$pulsar/magnetar/main.html.}).
However, without an estimate of $\dot{P}$, and thus $B$, the AXP
identification could not be confirmed, so further attempts were made
to re-detect the pulsar and pulsations.  Unfortunately, it was not
seen in a 1997 observation from the \textit{ASCA} Galactic Plane
Survey \citep{tkk+98}, and a 1999 pointed follow-up observation with
\textit{ASCA} found a possible counterpart, \axj\footnote{The position
of \axj\ has been corrected since the publication of \cite{vgtg00};
see \cite{tkgg06} for the best position.}, that was almost 10 times
fainter, too faint for a measurement of pulsations or a spectrum
\citep{vgtg00}.  \textit{Chandra}, \textit{XMM-Newton} and
\textit{BeppoSAX} observations during 2001-2003 revealed a point
source coincident with \axj\ and similar in brightness, but with a
slightly harder absorbed spectrum ($kT\sim1.0$~keV) than that seen for
\psr\ in 1993 \citep{isc+04}.

Presented here are the results of a \textit{Chandra X-ray Observatory}
monitoring campaign, conducted in 2003, with the goal of characterizing
the spectral and timing properties of \psr\ in a post-outburst state.

\section{\textit{Chandra} Observations}
\label{sec:obs}
Between June and September 2003, we obtained seven observations with
\textit{Chandra} ACIS-S in timed exposure mode.  The first six were
taken in 1/8 subarray mode in order to achieve high time resolution
(0.4~s); the seventh observation was full field.  Since the 1993
\textit{ASCA} position of \psr\ had a large (3$'$ radius) uncertainty,
we centered our observations at the \textit{Chandra} HRC position of a
possible counterpart (G. Israel, private communication).  All data
processing was performed using the CIAO 3.2.2 and CALDB 3.0.3 software
packages.

\begin{figure}
\centering
\includegraphics[width=0.48\textwidth]{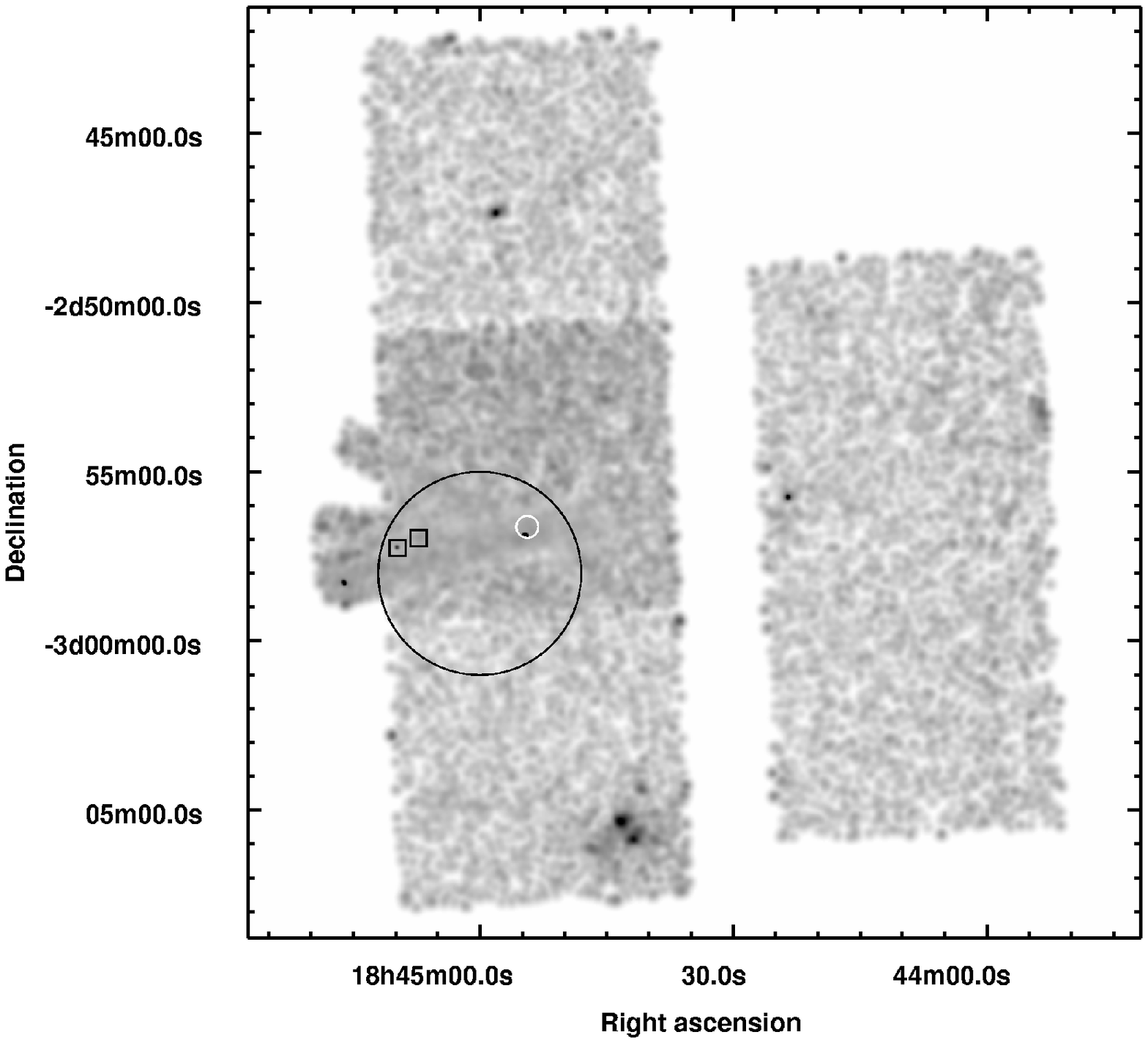}
\caption{Combined 2$-$10~keV \textit{Chandra} ACIS-S image of the
  field surrounding \psr, originally published in \cite{tkgg06}.  The
  potential counterpart \cxou\ falls within the 1993 and 1999
  \textit{ASCA} error circles of \psr\ (black circle, \citealt{gv98})
  and \axj\ (white circle, \citealt{vgtg00}), respectively.  Also
  indicated are two fainter point sources, CXOU J184507.2--025657
  (right box) and CXOU J184509.7--025715 (left box).}
\label{fig:image}
\end{figure}

One bright point source in the 3$'$
\textit{ASCA} error circle was found and designated \cxou\ (see
Fig.~\ref{fig:image}).  This is likely the counterpart to \axj, and
possibly \psr.  There was no evidence of extended emission.  

Two additional fainter point sources were detected 
inside the 1993 error circle of \psr\ but outside the 1999 error
circle of \axj\ (see Fig.~\ref{fig:image}).  One of them,
CXOU J184507.2--025657, was found coincident with a bright
near-infrared object from the 2MASS All Sky Survey; this challenges an
AXP interpretation, since all confirmed near-IR counterparts to AXPs
are very faint ($K\sim20$ mag) and \psr\ is known to be highly
absorbed ($N_H \ge 6\times 10^{22}$~cm$^{-2}$, \citealt{gv98}).

\section{Timing and Spectral Analysis}

Consider two cases: 1) \cxou\ is the counterpart to \psr, and 2)
\cxou\ is unrelated to \psr.  The results of the following analysis
were originally published in \cite{tkgg06}.

\paragraph{Case 1: \cxou\ is the counterpart.} We extracted light
curves from \cxou\ at the highest possible time resolution (0.4 s for
six observations, 3.2 s for the seventh) from each data set, in three
energy ranges: 1$-$10, 1$-$3, and 3$-$10~keV.  A fast fourier
transform (FFT) was performed on
barycentered event data, however, no evidence for pulsations was seen
in any of the resulting power density spectra.  For the longest
observation and the frequency range 0.0880$-$0.1436~Hz, we find a 95\%
confidence upper limit on the pulsed amplitude of 80\% in the
1$-$10~keV range.

\begin{figure}
\centering
\includegraphics[width=0.32\textwidth,angle=-90]{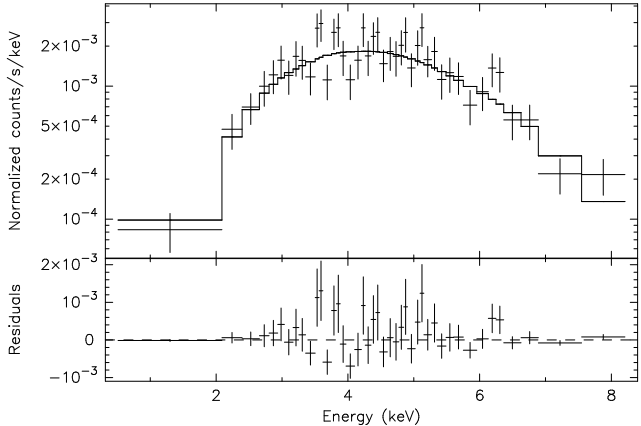}
\caption{The spectrum of \cxou\ shown with its best-fit blackbody
  model.}
\label{fig:spec}
\end{figure}

The individual observations contained insufficient counts to
adequately fit a spectrum, so we summed the extracted spectra into
one combined spectrum.  Using XSPEC 11.3.1, we found that the
background-subtracted combined spectrum was equally well
fit to a single-component absorbed thermal blackbody or power law:
Table~\ref{tab:spec} lists the best-fit spectral parameters, and 
Figure~\ref{fig:spec} shows the data fit to a blackbody.  Assuming
the blackbody model for now, we measured the combined
absorbed 2$-$10~keV flux to be $2.6 \pm 0.2 \times
10^{-13}$ \units; we also estimated the unabsorbed
flux to be $2.5-4.0\times 10^{-13}$ \units, taking
into account the 
large uncertainties on $N_H$.  Since the flux is roughly consistent
with that of \axj, we speculate that we have detected the same
object.
\begin{table}[t]
\caption{\cxou\ spectral parameters. Errors reflect 90\% confidence
  region. The absorbed flux is given for the 2$-$10~keV energy range,
  and we determine its uncertainty by fixing $N_H$ and $kT$ or
  $\Gamma$ at the best-fit value and adopting the fractional
  uncertainty on the normalization.}
\centering
\label{tab:spec}
\begin{tabular}{cccc}
\hline\noalign{\smallskip}
Model & $N_H$ & $kT$ (keV) & $F$\\[3pt]
 & (10$^{22}$ cm$^{-2}$) & or $\Gamma$ & (10$^{-13}$ \units)\\[3pt]
\tableheadseprule\noalign{\smallskip}
BB & $5.6^{+1.6}_{-1.2}$ & $2.0^{+0.4}_{-0.3}$ & $2.6\pm0.2$\\
PL & $7.8^{+2.3}_{-1.8}$ & $1.0^{+0.5}_{-0.3}$ & $2.8\pm0.2$\\
\noalign{\smallskip}\hline
\end{tabular}
\end{table}

Fixing $N_H$ and $kT$ at their best-fit blackbody values but allowing
the normalization to vary, we fitted the data from the seven
individual observations and found that the observed 2$-$10~keV flux at
each epoch was consistent with \cxou\ being constant over the 4-month
\textit{Chandra} observing period, at the combined flux value.  The
inset plot of Figure~\ref{fig:flux} shows this.

\paragraph{Case 2: \cxou\ is unrelated to \psr.}
We extracted light curves and spectra from the two
additional faint sources detected in the 3$'$ error region, but in all
instances there were not enough counts to detect pulsations or fit a
spectral model.  However, we noticed that most of the photons from
CXOU J184507.2--025657 were below 2~keV, which contradicts what is
known of \psr, namely that it is highly absorbed.  Because of
this and the aforementioned evidence in \S\ref{sec:obs}, we consider
CXOU J184507.2--025657 an unlikely counterpart to \psr.  For
CXOU J184509.7--025715, the data were insufficient for us to draw
meaningful conclusions about this source as candidate.

\psr\ may not have been re-detected at all, falling below the
3$\sigma$ background flux level.  We estimated an upper limit on a
hypothetical point source to be $\sim 8-13 \times 10^{-15}$ \units\ 
(2$-$10~keV), based on a variety of likely spectral models (see
Fig.~\ref{fig:flux}).

\section{A Transient AXP?}

\begin{figure}
\centering
  \includegraphics[width=0.48\textwidth]{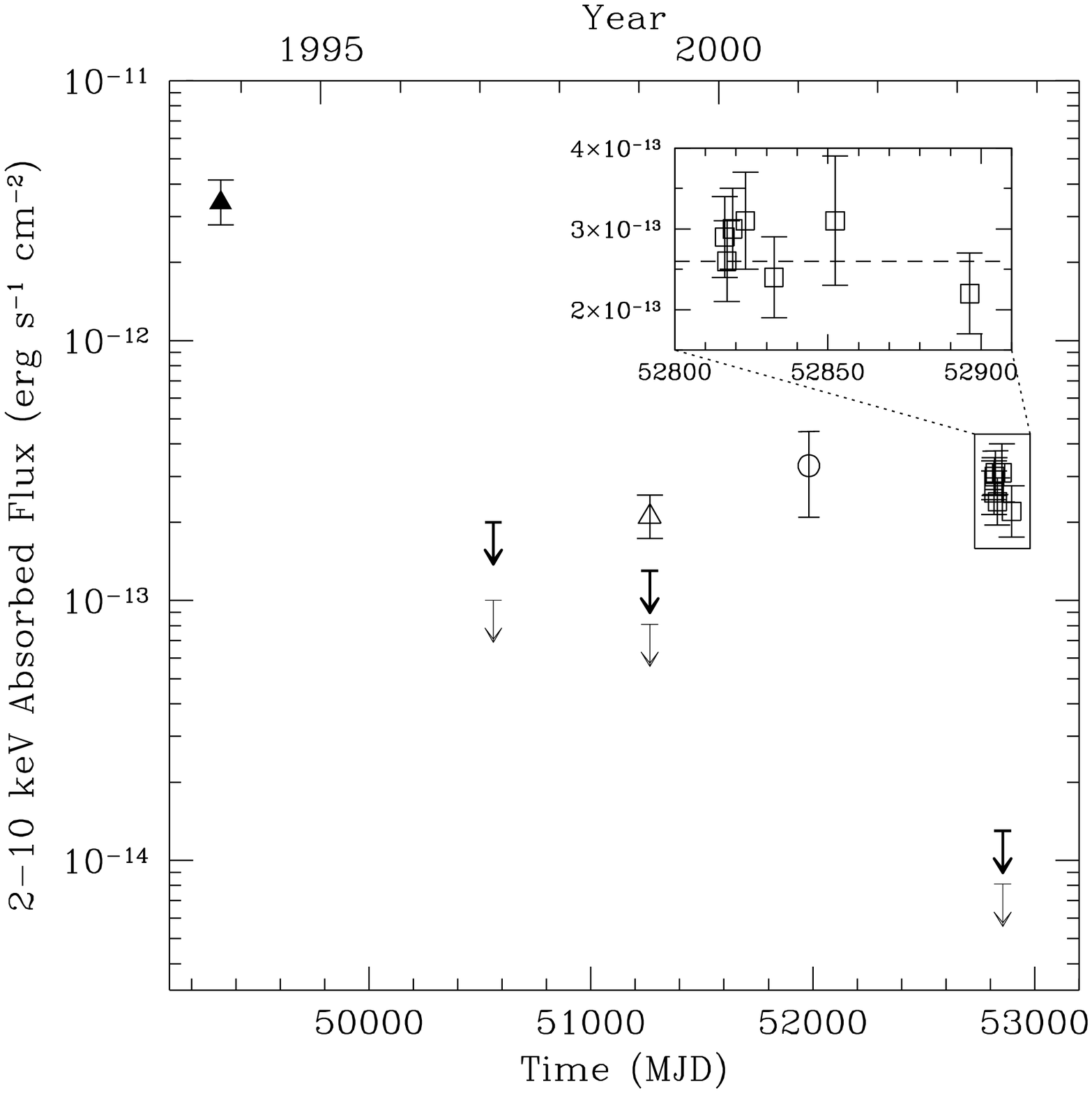}
\caption{The 10-year flux history of \psr, originally presented in
  \cite{tkgg06}.  The filled triangle
  is the original 1993 \textit{ASCA} detection of \psr; the open
  triangle is the 1999 \textit{ASCA} detection of \axj, and assumes
  the 1993 outburst spectrum.  The circle is the \textit{BeppoSAX}
  detection of
  a possible counterpart, and assumes the spectrum given by
  \cite{isc+04}.  Squares indicate the \textit{Chandra} detections of
  \cxou\ reported here, and assume the best-fit blackbody spectrum.
  We also represent the observed background levels as upper limits--in
  case the detections made were of unrelated objects--that assume the
  spectrum of \psr\ in outburst (thick arrows) and \xte\ in quiescence
  (thin arrows).  The \textit{Chandra} points are magnified in the
  inset plot, where the flux measured from the combined data set is
  indicated by the dashed line.}
\label{fig:flux} 
\end{figure}

Whether \cxou\ is truly the counterpart or not, its flux in 2003 is a
factor of $\sim$13 smaller than \psr's in 1993.  However if \psr\ has
not been re-detected at all, then this factor grows significantly
larger to $\sim$260$-$430, respresenting an unprecedented range in
variability for AXPs.  Figure~\ref{fig:flux} outlines the 10-year flux
evolution of \psr\ and its potential counterparts.

Comparable flux variability on large time scales has been seen in at
least one other AXP.  The 5.5-s transient AXP (TAXP) \xte\ was also
discovered when it was in a high state, in 2003 \citep{ims+04}, and
has since faded back towards its ``quiescent'' flux level
\citep{gh05}, as measured in archival \textit{ROSAT} observations from
1993 \citep{ghbb04}.  The pre-outburst source flux, which is nearly 2
orders of magnitude lower than its peak outburst flux, is much fainter
than that of any non-transient AXPs, bringing to mind the question of
how many more TAXPs have gone undetected in the Galaxy.

TAXPs are not accounted for in the framework of the magnetar model
\citep{td95,td96}, which attributes their persistent high-energy
emission to continual heating of and stresses on magnetar's crust.
The source of this crustal stress and heating is the gradual decay of
its ultra-high ($\sim$10$^{15}$~G) magnetic field, and can be used to
predict an X-ray luminosity that is well matched by that seen in
quiescent non-transient AXPs \citep{tlk02}.  X-ray bursts, like those
now observed in four AXPs including \xte\ \citep{wkg+05}, are thought
to result from the sudden fracturing of the magnetar's surface and
reconfiguring of its field lines.  So the question remains: if these
common elements link transient and non-transient AXPs as magnetars,
what is the cause of their differences?

Spectrally, we previously saw that \psr\ was not unlike other AXPs,
which typically have soft spectra (recall $kT\sim0.64$~keV during
outburst).  For this reason, the observed hardness of the 
\textit{Chandra} source ($kT\sim2.0$~keV) brings into question the
proposed association with \psr, and an overall AXP interpretation.
Moreover, \xte, the \textit{bona fide} TAXP, was observed to be harder
in outburst than quiescence ($kT\sim0.67$~keV compared to
$kT\sim0.18$~keV, respectively, from \citealt{ghbb04}), which is the 
opposite to what we have witnessed if \cxou\ is indeed a TAXP.

\section{Alternate Endings}
Given the uncertainty in the identity of \cxou, it seems prudent to
consider other plausible alternatives.  We argue on the basis of
key observable properties, such as its relatively hard spectrum,
intrinsic luminosity $L_X\approx 10^{33}(d/5)$~kpc, and apparent
stability on time scales of days to weeks.

\paragraph{Active galactic nuclei.}  The measured photon index
$\Gamma\sim 1.0$ from the power law model is not unlike that seen for
an active galactic nucleus (AGN, \citealt{woau04,nla+05}).
Using \textit{Chandra} ACIS-I, \cite{etp+05} studied the faint X-ray
emission from an ``empty'' Galactic plane region that was
conveniently centered only 1$^{\circ}$ away from our target, meaning
that they might have local properties in common such as $N_H$.  From
their models of Galactic source populations, we
estimate a $\sim$2\% likelihood that a circular region 3$'$ in radius
would contain a coincident AGN, $3\times 10^{-13}$ \units\
(2$-$10~keV) or
brighter.  Predicted optical/IR magnitudes fall at the limits of what 
current observatories are capable of, which will make it difficult to
conclusively confirm or rule out an AGN interpretation through such
means.

\paragraph{Galactic sources.}  Winds from massive stars have similar
spectral and flux properties, as do some high-mass X-ray binaries
\citep{mab+04}.  These systems, however, would tend to be bright in
optical/IR, which disagrees with the faint upper limit set by
\cite{isc+04} of $H>21$~mag.

Another group of Galactic objects with similar properties are
cataclysmic variables (CVs).  According to \cite{mab+04}, the IR
emission of CVs at comparable distances and extinctions to our
\textit{Chandra} source ought to be relatively faint, roughly
$K\approx22-25$~mag.  Therefore, it seems clear that optical/IR
observations alone will be insufficient to identify this source.

\section{Conclusions}

We have observed and analysed the \textit{Chandra} point source \cxou,
which may be the transient X-ray pulsar and candidate AXP \psr.
If it is the counterpart, then either \psr\ is not actually an
AXP, or AXPs are much more diverse in their spectral and flux
characteristics during quiescence than previously thought.  If it is
not the counterpart and \psr\ is an AXP, then the exhibited flux
variability presents a challenge to our current understanding of
AXPs as magnetars, and hints at a much larger population of faint AXPs
that remain undetected.

\begin{acknowledgements}
Thanks to G. Israel for providing details on the position
and flux of the possible counterpart to \psr. V.M.K. is a Canada
Research Chair and acknowledges funding from NSERC via a Discovery
Grant and Steacie Supplement, the FQRNT, and CIAR.  B.M.G. is an
Alfred P. Sloan Fellow and acknowledges support from Chandra GO grant
GO3-4089X, awarded by the SAO.
\end{acknowledgements}



\end{document}